\documentclass{article}%
\usepackage{graphicx}
\usepackage{amsmath}%
\usepackage{amsfonts}%
\usepackage{amssymb}
\begin{document}

\title{Brussels-Austin Nonequilibrium Statistical Mechanics in the Early Years:
Similarity Transformations between Deterministic and Probabilistic Descriptions}
\author{Robert C. Bishop$^{a,b}$\\$^{a}$Abteilung f\"{u}r Theorie und Datenanalyse, Institut\\f\"{u}r Grenzgebiete der Psychologie, Wilhelmstrasse\\3a, D-79098 Freiburg, Germany\\$^{b}$Permanent Address: Department of Philosophy,\\Logic and Scientific Method, The London School\\of Economics, Houghton St., London,\\WC2A 2AE, United Kingdom\\Draft}
\date{}
\maketitle

\begin{abstract}
The fundamental problem on which Ilya Prigogine and the Brussels-Austin Group
have focused can be stated briefly as follows. Our observations indicate that
there is an arrow of time in our experience of the world (e.g., decay of
unstable radioactive atoms like Uranium, or the mixing of cream in coffee).
Most of the fundamental equations of physics are time reversible, however,
presenting an apparent conflict between our theoretical descriptions and
experimental observations. Many have thought that the observed arrow of time
was either an artifact of our observations or due to very special initial
conditions. An alternative approach, followed by the Brussels-Austin Group, is
to consider the observed direction of time to be a basics physical phenomenon
and to develop a mathematical formalism that can describe this direction as
being due to the dynamics of physical systems. In part I of this essay, I
review and assess an attempt to carry out an approach that received much of
their attention from the early 1970s to the mid 1980s. In part II, I will
discuss their more recent approach using rigged Hilbert spaces.


Word count (including notes): 9,901

\end{abstract}

\noindent\pagebreak 

\begin{quotation}
How and to what extent the irreversible phenomena observed in the macroscopic
domain can be reconciled with the reversible dynamical laws of classical (or
quantum) mechanics is the fundamental question of statistical mechanics (Misra
1978, p. 1627).
\end{quotation}

\section{Introduction}

The work of Ilya Prigogine and his group is difficult to understand and
assess, being highly mathematical in nature. Moreover although their
fundamental intuitions have remained essentially unchanged over the course of
several decades, the approach has changed with time making their views
difficult to pin down with precision. The ideas Prigogine and his colleagues
have been pursuing in various forms were sketched in his (1962). He along with
George, Henin and Rosenfeld gave the earliest mathematically detailed
description in their (1973; see also George 1973a, 1973b).

The core idea is the following. The conventional approach to describing
physical systems within classical mechanics (CM) relies on a representation of
states $\omega$ (e.g. of particles) as points in an appropriate state space
$\Omega$. This means that the dynamics of a system are derivable from the
time-parameterized trajectories of these points. The equations governing the
dynamics of conservative systems are reversible with respect to time. When
there are too many states involved to make solving these equations feasible
(as in gases or liquids), coarse-grained averages--\textit{i.e.} macrolevel
averages ignoring microlevel details (so-called fine-grained level)--are used
to develop a statistical picture of how the system behaves rather than
focusing on the behavior of individual states. In contrast the Brussels-Austin
Group argues these systems should be approached in terms of models based on
distributions $\rho$ over an appropriate state space. These distribution
functions may be understood in terms of the probability density $\rho(\vec
{q}_{1},\vec{q}_{2},\vec{q}_{3},...,\vec{p}_{1},\vec{p}_{2},\vec{p}_{3}...,t)$
of finding a set of molecules (say) with coordinates $\vec{q}_{1},\vec{q}%
_{2},\vec{q}_{3},...$ and momenta $\vec{p}_{1},\vec{p}_{2},\vec{p}_{3}...$ at
time $t$ on the relevant energy surface and are analogous to the
microcanonical distribution. In the Brussels-Austin approach, the dynamics of
a system is calculated from \textit{distribution functions} directly. The
equations governing the dynamics of these distributions are generally
\textit{time-irreversible}. In addition interpreting the distribution
functions as probability densities suggests that macroscopic classical
statistical mechanics models are irreducibly probabilistic. This would mean
that probabilities are as much an ontologically fundamental element of the
macroscopic world, as successfully described by physics, as they are usually
taken to be for the microscopic world of quantum mechanics (QM).

The extent to which the Brussels-Austin program, as just sketched, is
distinguishable from a coarse-grained approach to statistical mechanics is a
delicate question. As this two-part essay develops, I will point out how their
work differs from typical coarse-grained approaches. My task in part I will be
to review Prigogine and co-workers' motivations (\S2) and then discuss their
various approaches to nonequilibrium systems, focusing on the period from the
1960s to the mid 1980s, covering their subdynamics and similarity
transformation approaches (\S\S3-4). Part II will focus on the more recent
rigged Hilbert space approach.

\section{Motivations}

A crucial motivation is the question of how classical dynamical systems,
described in conventional CM by deterministic, time-reversible equations of
motion, are related to time-irreversible processes. The central questions are:
What connections, if any, exist between these two types of systems; and Why is
it we never observe such processes going ``in reverse?''

A key concern is the status of the second law of thermodynamics: When a
constraint internal to a closed system is removed, the total entropy must
increase or at best stay constant. As one of the fundamental laws of
conventional equilibrium thermodynamics, it is valid only at or
near\textit{\ thermodynamic equilibrium}. Prigogine and his colleagues believe
that some appropriate generalization of the second law should be applicable to
nonequilibrium systems as well.

Typically coarse-grained descriptions of systems involve calculating
macro-level averages of quantities over finite volumes and are considered to
provide less specific information than descriptions involving points in state
space. Probabilistic processes in such models are irreversible, but are
usually interpreted as reducible; that is to say, the probabilities are
considered to be consequences of our calculation techniques and measurement
limitations. Irreversibility could then be understood as a consequence of a
coarse-grained description rather than as a fundamental feature of systems.

The second law is considered time-irreversible in so far as the process of
entropy increase cannot be reversed (a cube of ice melting and diffusing in a
glass of tea does not reconstitute itself into a cube of ice again). If the
second law is taken to be a fundamental law (not a consequence of
coarse-grained descriptions), then there is a puzzling conflict with our
fundamental time-symmetric equations, a particularly sharp conflict in the
case of conservative systems. Thus,

\begin{quotation}
[T]he elucidation of the relation between conservative and dissipative
dynamical systems necessarily involves a clarification of the relation between
deterministic dynamics and probabilities. Because of the close relation that
exists between entropy and probability, once this is clarified the relation
that exists between dynamics and the second law will also be made clear
(Nicolis and Prigogine 1989, p. 199).

In all fundamental theories (be it classical dynamics, quantum mechanics or
relativity theory) entropy is conserved as a result of the unitary (or
measure-preserving) character of the evolution, in flagrant contradiction with
the formulation of the second law of thermodynamics. As a result, the second
law has usually been regarded as an approximation or even as being subjective
in character. By contrast, in the approach to the problem of irreversibility
developed by us, the law of entropy increase and, therefore, the existence of
an ``arrow of time'' is taken to be a fundamental fact. The task of a
satisfactory theory of irreversibility is thus conceived as the study of the
fundamental change in the conceptual structure of dynamics, which the law of
entropy increase implies (Misra and Prigogine 1983, p. 421).
\end{quotation}

We can distinguish two types of irreversibility: \textit{extrinsic} and
\textit{intrinsic} (e.g. Atmanspacher, Bishop and Amann 2002). Extrinsic
irreversibility is irreversible behavior of a physical system due to its
interaction with an environment, where in the absence of an environment, the
system itself would be reversible. Examples of extrinsic irreversibility are
given by any open-system evolution described by a master equation. By
contrast, intrinsic irreversibility refers to irreversible behavior
originating in the dynamics of a physical system without explicit reference to
an environment. An example of intrinsic irreversibility would be kaon decay.
In contrast to most views on statistical mechanics (SM), the Brussels-Austin
Group believes that intrinsic irreversibility is fundamental and has been
searching for an intrinsically irreversible formulation of SM.\footnote{Their
original focus was on open systems. More recently they have focused on closed
systems (Part II).}

\section{Subdynamics and Similarity Transformations}

\subsection{Koopman Formulation of Classical Mechanics and its Extension}

Most of the Brussels-Austin Group's results are developed in a Hilbert space
(HS) in part due to matters of elegance as well as out of a desire to unify CM
and QM within one formalism. They relied heavily upon Koopman's extension of
HS and linear transformations to the study of steady $n$-dimensional fluid
flow with positive density (Koopman 1931). Originally, Koopman studied the
dynamics of state space volumes, where the elements $\phi$ of $L^{\infty
}(\Omega)$ are defined on the state space $\Omega$ (the natural dual space
being $L^{1}(\Omega)$). In CM, physical observables (e.g. energy) are usually
associated with real-valued functions defined on $\Omega$. Koopman's
formulation generalizes this feature in analogy with QM by associating linear
operators on $L^{\infty}(\Omega)$ with physical observables. Thus the
expectation value of an observable corresponding to the operator $F$ is
defined as $(\phi,F\phi)$, where $F(\phi)(\omega)=f(\omega)\phi(\omega)$ is
unitary and $f:\Omega\rightarrow\mathcal{R}$ is defined on state space points
$\omega\in\Omega$ referring to states.

An important relevant example is Liouville's equation describing the evolution
of a distribution function $\rho$ in state space. Liouville's equation can be
converted to an operator equation with the unitary\footnote{Koopman uses the
Hilbert space generator representation for $U_{t}$ which is formulated for
elements of $L^{2}$. This has led to some confusion as to what mathematical
spaces to which Koopman's formulation is applicable.} form $U_{t}=e^{-iLt}$
and, in Hamiltonian form,%
\begin{equation}
L=\sum_{k}\left[  \frac{\partial H}{\partial p_{k}}\frac{\partial}{\partial
q_{k}}-\frac{\partial H}{\partial q_{k}}\frac{\partial}{\partial p_{k}%
}\right]  \text{,}%
\end{equation}
where (1) is the Poisson bracket,\ $p$ and $q$ representing generalized
momenta and positions respectively, and $t$ representing time. The operator
$U_{t}$ is self-adjoint and is generally unbounded on $L^{\infty}(\Omega)$.
The physical interpretation of the Koopman formulation of Liouville's equation
is the same as in CM. In the so-called Lagrangian picture, the motion of state
space points can be treated like a flowing fluid. In both conventional SM and
the modified Koopman formulation, Liouville's equation describes the evolution
of the distribution $\rho$ as it moves through state space from the
perspective of an observer at rest.

Koopman's formalism is defined on the state space $L^{\infty}(\Omega)$. The
Brussels-Austin Group wanted to use this formalism to study the evolution of
state space points themselves and, so, extended Koopman's formalism to the
space $L_{%
%TCIMACRO{\U{b5}}%
%BeginExpansion
\mu
%EndExpansion
}^{2}(\Omega)$ (square integrable functions using measure $\mu$; e.g. Misra
1978), although no formal justification for these extensions was ever worked
out rigorously.

\subsection{Subdynamics}

The first approach developed in this early phase was called
\textit{subdynamics}, the idea being to split the state space $L_{%
%TCIMACRO{\U{b5}}%
%BeginExpansion
\mu
%EndExpansion
}^{2}(\Omega)$ of the system dynamics into distinct thermodynamic and
non-thermodynamic subspaces \textit{via} an appropriate projection operator,
and then to enumerate the conditions under which the non-thermodynamic
subspace made no contribution to the evolution of the thermodynamic features
of the system (Prigogine, George and Henin 1969; Prigogine et al. 1973;
Obcemea and Br\"{a}ndas 1983; Dougherty 1993; Karakostas 1996). Karakostas
(1996, pp. 383-4) argues that the 1973 version of subdynamics represents a
generalization of coarse-graining, in that it merely amounts to a reduced
description of the system.\footnote{Versions of subdynamics derived from a
Lyapunov variable are not so easily classified as coarse-grainings (see
below).} Ultimately, however, subdynamics turned out to be dependent on the
Brussels-Austin conception of the relationship between deterministic dynamics
and probabilistic dynamics--e.g. similarity transformations--so I will not say
anything more about subdynamics here.

\subsection{Similarity Transformations}

The second approach developed during this period was based on a similarity
transformation $\Lambda$ mapping a trajectory description of \ ``unstable''
classical systems--systems exhibiting exponential trajectory divergence--into
a description in terms of probabilistic Markov processes. The existence of
such a $\Lambda$ would then provide a means of translating between the
trajectory and the Markov descriptions. In a problematic sense to be discussed
below, this would establish an ``equivalence'' between trajectory and
probabilistic descriptions and, hence, an equivalence between time-reversible
and time-irreversible dynamics for such systems. Furthermore, although the
amount of information in the trajectory description is supposedly preserved in
moving to the probabilistic description, Prigogine and coworkers also claimed
that there was new physics contained in the latter description that was not
contained in the former. It is this additional physics in the probabilistic
description that they believed resulted in new physical features, rendered
elements of the trajectory description (e.g. ``exact trajectories'')
unphysical idealizations.

The technical details may be summarized as follows: When mathematically
defined on $L_{%
%TCIMACRO{\U{b5}}%
%BeginExpansion
\mu
%EndExpansion
}^{2}(\Omega)$, the evolution of particular types of Markov processes can be
shown to correspond to nonunitary semigroup operators $W_{t}^{\ast}$ =
$\Lambda U_{t}\Lambda^{-1}$ \textit{via} a similarity transformation
$\Lambda:L_{%
%TCIMACRO{\U{b5}}%
%BeginExpansion
\mu
%EndExpansion
}^{2}(\Omega)\rightarrow L_{%
%TCIMACRO{\U{b5}}%
%BeginExpansion
\mu
%EndExpansion
}^{2}(\Omega)$, where $\Lambda$ is closed, densely defined on $L_{%
%TCIMACRO{\U{b5}}%
%BeginExpansion
\mu
%EndExpansion
}^{2}(\Omega)$ and invertible.\footnote{Karakostas (1996) gives a detailed
discussion of these and related operators defined in the subdynamics approach
developed in (Prigogine \textit{et al}. 1973). The transformation discussed in
Karakostas (1996) is related to, but different from the similarity
transformation I discuss here as Prigogine and coworkers have made several
modifications to their program since publication of (Prigogine \textit{et al}.
1973). For example $\Lambda$ in this latter publication is star-unitary, but
the $\Lambda$ developed later in the approach I am reviewing is nonunitary.}
The crucial result is that$\ W_{t}^{\ast}$ be positivity preserving on the
positive $t$-axis only, guaranteeing that $W_{t}^{\ast}$ leads to a monotonic,
time-irreversible approach to a unique final state (conjectured in (Misra,
Prigogine and Courbage 1979, p. 12) and proven in (Goodrich, Gustafson and
Misra 1980)).\footnote{The proof is not constructive, however, so $\Lambda$
must still be constructed for every system.} Any system characterized by
$\Lambda$ would then be asymptotically stable: Any initial state will evolve
irreversibly to a unique equilibrium distribution as $t\rightarrow\infty$. By
contrast in Koopman's original formulation, dynamical systems are
characterized by a unitary group $U_{t}$ defined for both the positive and
negative $t$-axes, preserving the time reversibility of the governing
dynamical equations.\footnote{Originally the Brussels-Austin Group treated the
operators $\Lambda$ and $W_{t}^{\ast}$, along with the distribution $\rho$, as
being defined on SHS. This turns out to be inadequate, however, as I will
explain below.}

By constructing a nonunitary similarity transformation $\Lambda$ acting on the
distribution function $\rho$ in the trajectory description defined on $L_{%
%TCIMACRO{\U{b5}}%
%BeginExpansion
\mu
%EndExpansion
}^{2}(\Omega)$ at time $t$, a distribution function in the Markov description
is then given by $\rho^{\prime}=\Lambda\rho$. The time-reversible Liouville
equation in Koopman's original formulation,%
\begin{equation}
i\frac{\partial\rho}{\partial t}=U_{t}\rho\text{, }-\infty\leq t\leq
\infty\text{ ,}%
\end{equation}
where $\rho$ and $U_{t}$ are defined on $L_{%
%TCIMACRO{\U{b5}}%
%BeginExpansion
\mu
%EndExpansion
}^{2}(\Omega)$, is then transformed into a time-irreversible equation%
\begin{equation}
i\frac{\partial\rho^{\prime}}{\partial t}=W_{t}^{\ast}\rho^{\prime}\text{,
}t\geq0
\end{equation}
where $W_{t}^{\ast}$ and $\rho^{\prime}$ are also defined on $L_{%
%TCIMACRO{\U{b5}}%
%BeginExpansion
\mu
%EndExpansion
}^{2}(\Omega)$. The interpretation of (3) is similar to that of the Liouville
equation in the Koopman description: It describes the evolution of the density
function $\rho^{\prime}$ in state space, but only for the positive time
direction. The dynamics under $U_{t}$ is time-reversible. However in (3), the
evolution governed by $W_{t}^{\ast}$ is \textit{time-irreversible}. This is
the key for time-irreversibility in the similarity transformation approach and
leads to the definition of intrinsic randomness: A model is intrinsically
random if there exists a nonunitary $\Lambda$ such that the unitary group
$U_{t}$ is transformed to the Markov semigroup $W_{t}^{\ast}$ (Goldstein,
Misra and Courbage 1981, pp. 114-8; Courbage and Prigogine 1983, p. 2412).

The strategy was to use a nonunitary similarity transformation $\Lambda$ to
move from the trajectory description characterized by $U_{t}$ to the Markov
description characterized by $W_{t}^{\ast}$. Since similarity transformations
preserve all structural features, the hope was that the two descriptions would
be shown to be ''equivalent'' \textit{via} $\Lambda$.

\subsection{Microentropy Operator}

Following a suggestion by Misra (1978), $\Lambda$ was derived from the
so-called \textit{microentropy operator} $M$, a positive linear operator
defined on $L_{%
%TCIMACRO{\U{b5}}%
%BeginExpansion
\mu
%EndExpansion
}^{2}(\Omega)$ that, according to Misra, fulfils the conditions for a
\textit{Lyapunov variable}, i.e., a variable that increases monotonically to
an asymptotically stable value (Misra 1978; e.g. Hale and Ko\c{c}ak 1991, pp.
277-92).\footnote{That $M$ be a Lyapunov variable is important for a
generalization of the concept of entropy discussed below.} A system must have
at least the property of strong mixing in order for Lyapunov variables to
exist. Lyapunov variables can be formally constructed for Kolmogorov or
K-flows (1978, pp. 1629-30). Strong mixing is a necessary condition, while
being a K-flow is a sufficient condition for the existence of such variables.
For unstable systems Misra's proposed that $M$ be identified with an
appropriate Lyapunov variable. Later Gustafson showed that the $\Lambda$
transformation so defined exists only for K-flows (1997, pp. 61-4).

The operator $M$ obeys the following properties (Misra 1978; Braunss 1984):

\begin{enumerate}
\item[(i)] If $\rho\in L_{%
%TCIMACRO{\U{b5}}%
%BeginExpansion
\mu
%EndExpansion
}^{2}(\Omega)$, then $M\rho\in L_{%
%TCIMACRO{\U{b5}}%
%BeginExpansion
\mu
%EndExpansion
}^{2}(\Omega)$.

\item[(ii)] $M$ is nonnegative; that is, $(\rho_{t},M\rho_{t})>0$ for all
$t\geq0$ and decreases monotonically to a minimum value for the equilibrium
distribution $\rho_{eq}$, where $\rho_{t}=U_{t}\rho$.

\item[(iii)] $d/dt(\rho_{t},M\rho_{t})\leq0$ for all $t\geq0$.
\end{enumerate}

\noindent Property (i) expresses closure: $M$ never leads outside $L_{%
%TCIMACRO{\U{b5}}%
%BeginExpansion
\mu
%EndExpansion
}^{2}(\Omega)$. Properties (ii) and (iii) characterize a Lyapunov variable,
implying that $M$ is monotonically increasing. Since $M$ is positive, it can
be factorized (\textit{i.e.} $M=\Lambda^{\ast}\Lambda$), so $\Lambda=$ $M^{%
%TCIMACRO{\U{bd}}%
%BeginExpansion
\frac12
%EndExpansion
}$. Furthermore as Braunss pointed out, the microentropy operator illuminates
the Brussels-Austin definition of intrinsic randomness. Suppose the dynamics
for a K-flow in the trajectory description is given by $U_{t}$ and that $M$ is
a Lyapunov variable for this dynamics. Then $(\rho_{t},M\rho_{t})^{1/2}$
defines a contractive semigroup $W_{t}^{\ast}$ (namely $W_{t}^{\ast}$ $=$
$\Lambda U_{t}\Lambda^{-1}$), that can act to convert smooth, Hamiltonian
trajectories into Brownian trajectories (1985, p. 9-11).

The factorization of $M$ is not unique, however, because in general operators
$\Lambda$ differ by phase factors, where $W_{t}^{\ast}$ $\varphi_{\Lambda
}|\Lambda|=$ $|\Lambda|U_{t}$ and $\varphi_{\Lambda}$ is the phase factor for
$\Lambda$. So the class of $\Lambda$ transformations must be restricted to
\textit{admissible factorizations}, i.e. where $W_{t}^{\ast}\Lambda=\Lambda
U_{t}$ holds (Braunss 1985, 19). Furthermore, there is a practical difficulty
in identifying an appropriate positive definite variable serving as a basis
for $M$, there being no constructive guidance for choosing appropriate variables.

Misra's proposal of relating $\Lambda$ to $M$ also allows $\Lambda$ to be
related to a time operator $T$ for K-flows (Misra 1978; Misra, Prigogine and
Courbage 1979; Goldstein, Misra and Courbage 1981). Let $\mathcal{H}_{0}$
denote the one-dimensional subspace spanned by constant-valued functions on a
given energy surface, $P_{0}$ the projections from $L_{%
%TCIMACRO{\U{b5}}%
%BeginExpansion
\mu
%EndExpansion
}^{2}(\Omega)$ onto this subspace, $\mathcal{H}_{0}^{\perp}$ the orthogonal
complement of $\mathcal{H}_{0}$ and $P_{0}^{\perp}$ the projections from $L_{%
%TCIMACRO{\U{b5}}%
%BeginExpansion
\mu
%EndExpansion
}^{2}(\Omega)$ onto $\mathcal{H}_{0}^{\perp}$. For dynamical systems that are
K-flows, there exists a family of projection operators $F_{\eta}$,
$-\infty<\eta<\infty$, with the following properties (Misra 1978, p. 1629):

\begin{enumerate}
\item[(i)] $F_{\eta}\leq F_{\kappa}$ if $\eta<\kappa$.

\item[(ii)] $\lim_{\eta\longrightarrow\infty}F_{\eta}=P_{0}^{\perp}$.

\item[(iii)] $\lim_{\eta\longrightarrow-\infty}F_{\eta}=0$.

\item[(iv)] $F_{\eta}$ is strongly continuous in $\eta$.

\item[(v)] $U_{t}F_{\eta}U_{t}^{\dagger}=F_{\eta+t}$.
\end{enumerate}

\noindent Conditions (ii) and (iii) are to be understood in the strong
operator limit. It is then possible to construct a self-adjoint operator%
\begin{equation}
T=\int_{-\infty}^{\infty}\eta dF_{\eta}\text{,}%
\end{equation}
which has $F_{\eta}$ as its spectral family of projections. $T$ is
self-adjoint and canonically conjugate to $L$ \textit{via} the commutation
relation $[L,T]=-iI$, where $I$ is the identity operator and the eigenvalues
of $T$ are determined by the parameter time $t$ when applied to eigenfunctions
of $T$. Then $M$ can be constructed as a function of $T$ as%
\begin{equation}
M=h(T)+\alpha P_{0}\text{,}%
\end{equation}
where $\alpha\geq0$ and $h(T)$ is an operator defined through the relation%
\begin{equation}
(\psi,h(T)\phi)=\int_{-\infty}^{\infty}h(\eta)d(\psi,F_{\eta}\phi)\text{,}%
\end{equation}
where $\psi$, $\phi$ are vectors in $\mathcal{H}_{0}^{\perp}$ and $h(\eta)$ is
any monotone decreasing, positive, bounded, continuous and differentiable
function, whose derivative is always negative and bounded (Misra 1978, p.
1629). The inner product in (6) is to be understood in the following way. To
each vector $\psi$ in $\mathcal{H}_{0}^{\perp}$, there is a corresponding
family of functions $\{\psi_{n}(\eta)\}$, where $n=1,2,3,...$ and where
$\psi_{n}(\eta)\in L_{%
%TCIMACRO{\U{b5}}%
%BeginExpansion
\mu
%EndExpansion
}^{2}$, such that%
\begin{equation}
(\psi,\phi)=\sum_{n=1}^{\infty}\left(  \int_{-\infty}^{\infty}\psi_{n}^{\ast
}(\eta)\phi(\eta)d\eta\right)  .
\end{equation}
The operator $M$ as defined in (5) satisfies all the properties listed above.

The time operator $T$ associates an ``age'' or ``internal time'' with a well
defined distribution function $\bar{\rho}\equiv\rho-\rho_{eq}$. If $\bar{\rho
}$ is an eigenfunction of $T$, the corresponding eigenvalue gives the ``age''
associated with $\rho$ and is determined by an external (i.e. observer's) time
that serves to label the dynamics. If $\bar{\rho}$ is not an eigenfunction of
$T$, but a combination of eigenfunctions corresponding to two or more distinct
eigenvalues, then only an ``average age'' can be associated with $\rho$
(Misra, Prigogine and Courbage 1979, pp. 17-8).

The discovery of a time operator was one of the Brussels-Austin Group's
significant contributions to SM and systems theory (c.f. Atmanspacher and
Scheingraber 1987, where an alternative derivation of a time operator is given
and compared with the Brussels-Austin version). This ``internal time''
operator, according to Misra, Prigogine and Courbage, expresses the `inherent
(but hidden) stochastic and nondeterministic character of the evolution' of
such unstable systems (1979, p.5). It is this hidden `stochastic and
nondeterministic' character of the evolution that the change of representation
\textit{via} $\Lambda$ is supposed to reveal.

Furthermore, the existence of $\Lambda$\ leads to the claim that for unstable
CM systems, one can find a representation in which the dynamics are
irreducibly probabilistic in that $\Lambda$ transforms a trajectory
representation into a probabilistic one, so that unstable classical systems do
not possess exact smooth (i.e., everywhere differentiable) trajectories in the
probabilistic description (see \S3.5 below).

In addition to finding a time-irreversible dynamics, Prigogine and co-workers
were also interested in far from equilibrium SM. In conventional equilibrium
SM, the concept of entropy is defined at equilibrium. For example starting
with the canonical probability distribution, the fine-grained Gibbsian entropy
may be expressed as%
\begin{equation}
-\int_{\Omega}\rho\ln\rho d\Omega\text{ .}%
\end{equation}
As a first step toward a conception of entropy valid for nonequilibrium as
well as equilibrium cases, Misra proposed a generalization of the conventional
definition (8),%
\begin{equation}
-\ln(\rho,M\rho)\text{ ,}%
\end{equation}
where $\psi$ represents normalized functions on $L_{%
%TCIMACRO{\U{b5}}%
%BeginExpansion
\mu
%EndExpansion
}^{2}(\Omega)$ and%
\begin{equation}
\rho(\omega)=|\psi(\omega)|^{2}\text{,}%
\end{equation}
where (10) can be interpreted as the Gibbsian ensemble. Since near equilibrium
the thermodynamic entropy of a system increases monotonically until
equilibrium is reached, the idea is that any function used to model
thermodynamic entropy far from equilibrium should also have the property of
increasing monotonically in the neighborhood of a nonequilibrium stable
state.\footnote{For example, (8) is not monotonically increasing for
Hamiltonian models (in fact it is an invariant of Hamiltonian evolution). A
coarse-grained average version of (8) combined with some probabilistic
assumptions yields a monotonically increasing function for Hamiltonian models
(Misra 1978, p. 1630). The proposal in (9) is monotonically increasing in
these models without resorting to such assumptions.} Misra's suggestion is
that functionals like (9) may be interpreted as a nonequilibrium entropy since
they monotonically increase with time (1978, p. 1627).\footnote{Atmanspacher
and Scheingraber (1987) proposes characterizing nonequilibrium systems without
utilizing the concept of entropy.}

The quantity%
\begin{equation}
(\psi,F\psi)=\int_{\Omega}f(\omega)\rho(\omega)d\mu(\Omega)
\end{equation}
can be interpreted as an expectation value of the observable $F$ in the state
$\psi$, where $F$ is the operator of multiplication by the state space
function $f(\omega)$. For observables corresponding to state space functions
$f(\omega)$, the expression (11) represents an ensemble average of $f(\omega)$
in the ensemble (10). The correspondence between functions $F$ and the
ensembles $\rho(\omega)$ is one-to-one because $(\psi,F\psi)$ is single-valued
when $F$ is an operator of multiplication on state space (Misra 1978, p.
1628). By contrast the functional $(\psi,M\psi)$ is \textit{not} a
single-valued functional of $\rho(\omega)$, though it may be possible to
ensure that the rate of change of such a functional remains a single-valued
functional of $\rho(\omega)$ (Misra 1978, p. 1628). A further consequence is
that Lyapunov variables such as $M$ must fail to commute with at least some of
the operators that are multiplications by functions defined on state space
(Misra 1978, p. 1628). This implies that unstable systems possess noncommuting
observables (see below).

To summarize, according to the Brussels-Austin Group, provided such a
$\Lambda$ can be found, the deterministic dynamics of unstable systems are
equivalent to time-irreversible, (irreducibly) probabilistic Markov processes.
The first concrete example they gave was for a simplified Baker's
transformation.\footnote{The Baker's transformation is discussed in more
detail in \S3.5.} Although it is conservative and time reversible, the
question naturally arises: Are there any realistic physical systems for which
$\Lambda$ could be constructed? Generic prescriptions for constructing time
operators for Bernoulli systems and Kolmogorov flows were given in (Courbage
and Misra 1980; Goldstein, Misra and Courbage 1981). These constructions,
however, are purely formal and, hence, $\Lambda$ also remains
formal.\footnote{An alternative derivation of $\Lambda$ and its relationship
to time operators and the evolution of states is given in (Suchanecki 1992).}

More recently, concrete examples of time operators have been constructed for
unilateral shift representations of dynamics for Renyi maps, where $T$ for
such maps is densely defined on a HS (Antoniou, Sadovnichii and Shkarin 1999;
Antoniou and Suchanecki 2000; Mercik and Weron 2000). Furthermore, a generic
prescription for constructing $T$ for exact systems, where the dynamics is
noninvertible, was given in (Antoniou and Suchanecki 2000). Time operators for
the semigroups associated with the nonrelativistic and relativistic
one-dimensional diffusion equations have also been constructed (Antoniou,
Prigogine, Sadovnichii and Shkarin 2000). Nevertheless, in all these cases
$\Lambda$ appears only as a formal object assumed to be densely defined on a HS.

\subsection{Trajectories}

Another of the Brussels-Austin Group's claims is that the exact deterministic
trajectories of unstable dynamical systems are \textit{idealizations}. There
has been a great deal of confusion in understanding precisely what Prigogine
and collaborators have meant when they write that exact deterministic
trajectories do not exist or are unrealizable in unstable systems. Many have
interpreted their statements and arguments to mean the \textit{total absence}
of trajectories for such systems (e.g. Bricmont 1995). But the Brussels-Austin
Group only meant to be arguing against a \textit{particular type} of
trajectory, namely those which have unchanging width and are everywhere
differentiable (``exact and smooth'' in the Brussels-Austin nomenclature).
However, this distinction did not receive sufficient emphasis in the
similarity transformation approach and was easily misunderstood.

Misra, Prigogine and Courbage (1979) argued for the ``unreality'' of these
exact deterministic trajectories in the following way. In unstable systems,
viewed from a Lagrangian point of view, where the state space points are in
motion, small regions of state space contain points moving along `rapidly
diverging or qualitatively distinct types of trajectories.' The conclusion
they draw is that

\begin{quotation}
[o]bviously, in this situation, the concept of deterministic evolution along
state space trajectories cannot be defined operationally and hence,
constitutes a physically unrealizable idealization. Therefore, in dealing with
dynamically unstable systems, classical mechanics seems to have reached the
limit of the applicability of some of its own concepts. This limitation on the
applicability of the classical concept of state space trajectories is--it
seems to us--of a fundamental character. It forces upon us the necessity of a
new approach to the theory of dynamical evolution of such systems which
involves the use of distribution functions in an essential manner (Misra,
Prigogine and Courbage 1979, pp. 4-5).
\end{quotation}

Leaving aside the questionable association of physical processes with
operational definitions, there are two things to note about this passage.
First, at the time of the above quotation, the Brussels-Austin Group had not
given detailed arguments that the concept of exact, smooth trajectories was
physically unrealizable. Rather they viewed the failure of the concept as
``obvious'' for unstable systems, the classical concept of exact, smooth
trajectory having ``reached the limit of applicability'' in such systems.
Second, they took this failure to mean that such dynamical systems must be
described by distribution functions implying \textit{probabilistic
descriptions are fundamental for such systems}-- i.e., intrinsic randomness.

Their argument against the reality of exact, smooth trajectories later took
the following form:

\begin{enumerate}
\item[(A)] Deterministic dynamics and Markov processes are ``equivalent''
descriptions for unstable systems via the existence of the transformation
$\Lambda$.

\item[(B)] However, the concepts of point-like states and exact, smooth
trajectories are physically unrealizable idealizations for unstable systems.

\item[(C)] On the other hand, Markov processes are operationally well defined
for unstable systems.

\item[(D)] Therefore, evolution of probabilistic distributions, not state
space point trajectories, represent the fundamental descriptions of such systems.
\end{enumerate}

\noindent In (D) Prigogine and his colleagues are urging upon us a
fundamentally different way of conceiving classical unstable systems, not
merely a new formalism for calculating results on such systems. Premises (A)
and (B) are both required for (D). Premise (A) comes from the ``equivalence''
thesis between deterministic dynamics and probabilistic processes mentioned
above. The requirement of (B) is more subtle. Recall that the classical
conception of state space is a space of points each of which represents a
possible state of the system in terms of particular values of the positions
and momenta of the system constituents. The failure of the concept of smooth
deterministic trajectories implies that exact states for such systems are also
\textit{unphysical idealizations}, meaning probabilities can arise in some
fashion other than coarse-graining. The unreality of exact states means that
the state space points in CM are also unphysical idealizations: `The concept
of state space point and state space trajectories, which are regarded in the
classical theory as simple and basic notions, must be viewed now as a
mathematical reconstruction, and this reconstruction requires infinite
precision' (Misra and Prigogine 1983, p. 427; see also Goldstein, Misra and
Courbage 1981, pp. 112-3). Prigogine and coworkers considered infinite
precision to be physically impossible, hence their appeal to premise (B) as
well for the conclusion that probability is irreducible in unstable systems.

Given such far-reaching claims, the similarity transformation approach
warrants closer examination. I will first assess premises (A) and (B)
independently and then spell out what conclusions I believe the approach
licenses. Along the way, I will indicate revisions the Brussels-Austin Group
made to their approach.

Although some in the Brussels-Austin Group have taken it as ``obvious'' that
the concepts of exact states and smooth trajectories fail for unstable
systems, the failure of these concepts is not so obvious. Simply invoking
operationalism is no longer a convincing argument. This attempt to justify (B)
is particularly weak since it can at best mean that \textit{in practice} our
operational definition suffers an empirical breakdown for unstable systems.
Our usual procedures in classical mechanics will not allow us to predict with
accuracy the trajectories of the system arbitrarily far into the future
because of our inability to either measure or represent the initial conditions
to infinite accuracy (Bishop, forthcoming). One cannot conclude from this,
however, that exact trajectories and state space points do not exist.

A revised argument involves extending $\Lambda$ and $W_{t}^{\ast}$ to
generalized distribution functions from the theory of distributions (i.e.
using the theory of distributions; c.f. Schwartz 1950, 1951). Originally
$\Lambda$ and $W_{t}^{\ast}$ were supposedly defined as acting on distribution
functions $\rho$ defined on $L_{%
%TCIMACRO{\U{b5}}%
%BeginExpansion
\mu
%EndExpansion
}^{2}(\Omega)$. Misra and Prigogine (1983) claimed that since `we are
interested in studying the evolution (under [$W_{t}^{\ast}$]) of phase points,
we need to extend the action of $\Lambda$ and [$W_{t}^{\ast}$] to singular
(Dirac $\delta$-functions type) distributions concentrated on a given state
space point' (Misra and Prigogine 1983, pp. 423-5).\footnote{Mathematically
this extension requires that the Hilbert space be extended as well. Initially
they were unaware of this point, but extended spaces becomes important in
their more recent work discussed in Part II.} They explicitly constructed
$U_{t}$, $\Lambda$ and $W_{t}^{\ast}$ and a complete set of orthonormal
eigenvectors for $W_{t}^{\ast}$ for the Baker transformation. It turns out
that both regular and singular distribution functions can be expanded as
linear combinations of these eigenvectors. This allows the action of $\Lambda$
and $W_{t}^{\ast}$ to be applied to singular distributions like $\delta
(p-p_{_{0}})\delta(q-q_{_{0}})$, representing a distribution concentrated at
the point $(p_{_{0}},q_{_{0}})$ in state space. Applying $\Lambda$ to
$\delta(p-p_{_{0}})\delta(q-q_{_{0}})$ transforms it from a function taking a
nonzero value \textit{only} at the point $(p_{_{0}},q_{_{0}})$ to a function
taking nonzero values over a subset of state space points (Goldstein, Misra
and Courbage 1981, 121). Something similar happens under the action of
$W_{t}^{\ast}$. Misra and Prigogine pointed out `that \textit{even if one
could start} with an initial condition corresponding to a point on the state
space, it will cease to be a state space point under the physical evolution
[$W_{t}^{\ast}$] and the transformation $\Lambda$' (1983, 424). From these
results they concluded that

\begin{quotation}
The basic object of the theory must now be not the state space points and
their dynamical evolution along state space trajectories, but the
transformation of points under the transformation $\Lambda$ and their
evolution under [$W_{t}^{\ast}$]. One might still argue that at least in the
case when $\Lambda$ is invertible, one could reconstruct the motion of state
space points along trajectories from a knowledge of the evolution under
[$W_{t}^{\ast}$] of the transformed object [$\delta(p-p_{_{0}})\delta
(q-q_{_{0}})$] (Misra and Prigogine 1983, 425).
\end{quotation}

They go on to argue that such a reconstruction is not possible for arbitrarily
large time `except if one assumes infinite accuracy in the observation of the
physically evolving states' (p. 425). This is consistent with the fact that
the dynamics of state space points and trajectories are not the ``fundamental
objects'' of their physical theory. Rather, under the action of $\Lambda$, the
distributions are now fundamental.

However, Batterman pointed out that this confused the evolution of Dirac-type
functions with that of points in state space (1991, pp. 259-260). state space
points $\omega\in\Omega$ are not the same type of mathematical objects as
distributions. So nothing was actually demonstrated about the dynamics of
state space points. Indeed no such line of demonstration can work, because the
operator $\Lambda$ in these cases maps singular distributions defined
\textit{on} $\Omega$ into distributions with finite support on $\Omega$, so
the probabilistic description describes the evolution of distributions on
state space points and not the evolution of the points themselves.
Furthermore, technically $M$, $\Lambda$ and $W_{t}^{\ast}$ are defined not for
points, because the vectors of $L_{%
%TCIMACRO{\U{b5}}%
%BeginExpansion
\mu
%EndExpansion
}^{2}$ are not points, but equivalence classes of functions that are equal
almost everywhere. This is a crucial point for the Brussels-Austin Group
because state space points represent the exact states of the system. The
latter simply drop out of the description, implying nothing about the nature
of their trajectories.

Note as well that this arguments confuses an epistemological claim (i.e. the
inability to attain infinite measurement accuracy) with an ontological one
(i.e. the ultimate nature of trajectories), a conflation of epistemology with
ontology plaguing nearly every one of the group's arguments regarding
trajectories in their older approach.

Misra (1978, pp. 1628-9) hints at another possible argument in support of (B).
In the Koopman formalism, classical observables are often associated with
linear operators that are multiplications by state space functions at least
some of which, according to Misra (1978), fail to commute with the
microentropy operator $M$. Sufficiently unstable classical models, so the
argument goes, would possess complementary observables in analogy with the
position and momentum operators in QM. Therefore the simultaneous
determination of some classical observables and a nonequilibrium entropy of
the form (9) would be subject to a Heisenberg-like relation. Misra states that
for conditions where the dynamics are described in terms of smooth state space
trajectories with all classical observables completely determined, the concept
of nonequilibrium entropy would be inapplicable. Alternatively, conditions
permitting the precise determination of the nonequilibrium entropy would
preclude the possibility of accurately determining the state space
trajectories (Misra 1978, p. 1629). The deterministic description in terms of
trajectories would then be complementary to the probabilistic description in
terms of nonequilibrium entropy.

The argument needs to be spelled out in more detail. First it needs to be
demonstrated that in the extended Koopman formalism all the physically
relevant classical observables can be representable as multiplications by
state space functions. As Misra points out, there are many more operators in
$L_{%
%TCIMACRO{\U{b5}}%
%BeginExpansion
\mu
%EndExpansion
}^{2}$ that are not multiplications, but these other operators are considered
to have no physically meaningful interpretation in terms of classical
observable quantities (1978, p. 1628). The physically meaningful operator $M$,
however, cannot be represented as a multiplication operator on state space due
to the requirement that it be monotonically increasing (Misra 1978, pp.
1627-8), so additional argumentation is needed to show that all \textit{other}
relevant classical observables must be represented by multiplications on state
space. An additional complication is that although in the ``momentum
representation'' in the extended Koopman formalism, the momentum is
represented by a multiplication operator, in the ``position representation'',
momentum is represented by a differential operator. So the form of some
physically meaningful operators are representation dependent. Furthermore
viable candidates for the mathematical descriptions of the supposed
nonequilibrium entropy need to be constructed for any realistic physical
systems. That question is related to the truth of (A) insofar as $\Lambda$ is
a function of $M$.

It is possible to give a less radical reinterpretation of (D) based on an
operationalist view of unobservable entities implicit in (Nicolis and
Prigogine 1989):

\begin{enumerate}
\item[(D')] Probabilistic descriptions represent the proper descriptions
consistent with observability and computability.
\end{enumerate}

\noindent It then remains to be seen what can be made of the claim that the
probabilistic description represents no ``loss of information'' with respect
to the trajectory description. In this context I need to examine a more recent
version of the argument for \textit{indiscernibility} of trajectories put
forward by Nicolis and Prigogine (1989, pp. 204-8). This argument begins by
noting that the Baker's transformation,
\begin{subequations}
\begin{align}
x^{\prime}  &  =2x\text{, \ \ \ \ \ \ \ }y^{\prime}=\frac{y}{2}\text{
\ \ \ \ \ \ }0\leq x\leq\frac{1}{2}\\
x^{\prime}  &  =2x-1\text{, \ }y^{\prime}=\frac{y+1}{2}\text{ \ }\frac{1}%
{2}\leq x\leq1\text{,}%
\end{align}
\newline is measure-preserving in the sense that the unit square remains
invariant under repeated iterations. The action of (12) stretches the square
in the $x$-direction as much as it compresses it in the $y$-direction. This
contraction plays a crucial role in Nicolis and Prigogine's indiscernibility
argument regarding the supposed ``irrelevance'' of Liouville's theorem for a
given sub-area of the unit square.

Suppose $\rho$ is confined initially to a sub-area $\Delta$ of state space
(e.g. the shaded region in Figure 1). The sub-area $\Delta$ remains invariant
through the first few iterations. According to Nicolis and Prigogine, the
contracting dimension decreases exponentially to some $\varepsilon$ and
remains constant thereafter, because \textit{this represents the limit of our
ability to measure or localize this dimension accurately}. They refer to the
decreasing dimension as a fiber which eventually gets distributed evenly
throughout state space and cannot be localized. Hence the support of $\Delta$
would increase until it reaches the support of a larger subset of the state
space and, from an \textit{empirical} standpoint, render Liouville's theorem
on preservation of the area of sub-regions of state space ``irrelevant''.
Nicolis and Prigogine interpret this as due to the fact that measurements (or,
more generally, effects of physical interactions) refer to regions of finite
support in state space, not to mathematical points. Supposedly the conclusion
regarding the indiscernibility of contracting fibers (analogous to state space
trajectories) follows from the fact that measurements are only finitely
accurate. Hence the support of the subregion $\Delta$ is not area preserving
below the observational threshold $\varepsilon$, so they conclude that
Liouville's theorem is inapplicable below this threshold.
%TCIMACRO{\FRAME{ftbpFU}{4.4166in}{0.7731in}{0pt}{\Qcb{Figure 1.The first steps
%in the stretching and folding action of the Baker's transformation.}}%
%{}{baker.wpg}{\special{ language "Scientific Word";  type "GRAPHIC";
%width 4.4166in;  height 0.7731in;  depth 0pt;  original-width 4.3656in;
%original-height 0.742in;  cropleft "0";  croptop "1";  cropright "1";
%cropbottom "0";  filename 'Baker.wpg';file-properties "XNPEU";}} }%
%BeginExpansion
\begin{figure}
[ptb]
\begin{center}
\includegraphics[
natheight=0.742000in,
natwidth=4.365600in,
height=0.7731in,
width=4.4166in
]%
{Baker.wpg}%
\caption{Figure 1.The first steps in the stretching and folding action of the
Baker's transformation.}%
\end{center}
\end{figure}
%EndExpansion

Clearly the argument confuses ontology with epistemology. Obviously, given
(12), the contracting dimension will never reach a constant $\varepsilon$;
rather, the fiber will asymptotically approach zero thickness. Hence the
density of $\Delta$ will remain invariant as its support varies. The
``irrelevance'' of Liouville's theorem, according to Nicolis and Prigogine, is
due to our inability to localize/measure the fiber below $\varepsilon$
thickness. But this is an epistemological point as the fibers of the Baker's
transformation remain ontologically precise according to (12). Therefore,
while there may be some epistemic bound on our ability to precisely measure
fibers of decreasing thickness, such a bound would not support the ontological
claim (B).

The existence of this epistemic bound does imply a loss of information about
the trajectories, so how can Prigogine and coworkers claim that their approach
suffers no loss of information when it is unable to demonstrate the unreality
of exact state space points and smooth trajectories and, hence, support (A)?
This argument can only be seen as supporting (A) if one adopts an operational
attitude: Unobservable trajectories will have the same observational
consequences as if there were no exact smooth trajectories at all. Since the
observational outcomes are the same, the concepts of smooth trajectories and
exact states can be dropped from the theoretical description as so much excess
baggage. This operational attitude is the only consideration that can be
offered in support of (D) and, of course, is compatible with (D$^{\prime}$).

In my judgment, the ontological conclusion (D) is not licensed in the
similarity transformation approach. The Brussels-Austin Group formalism does
not show the nonexistence of exact states or smooth trajectories in unstable
systems.\footnote{As the Brussels-Austin Group shifted away from the
similarity transformation approach, they dropped terms like ``nonexistent'' in
favor of terms like ``irrelevant'' when discussing trajectories (Part II).}
Rather, like the coarse-grained approaches from which the Prigogine school
seeks to distinguish itself, their new formalism substitutes a probabilistic
description in place of point-like states and trajectories in state space. In
contrast their formalism emphasizes physics ignored in typical coarse-graining
techniques. Furthermore standard coarse-graining replaces individual states
$\omega$ with distributions defined uniformly over a finite cell of $\Omega$
without distinguishing points belonging to different stable manifolds in
unstable systems. The Brussels-Austin probability distributions are
constructed to distinguish points belonging to stable manifolds from unstable
ones. Furthermore the Markovian semigroups derived from $\Lambda$ are not
related to local point transformations in state space in contrast to those
semigroups derived from coarse-graining projections (Suchanecki, Antoniou and
Tasaki 1994). So the similarity transformation approach can be viewed as an
alternative calculational approach to coarse-graining provided $\Lambda$ can
be constructed for real-world systems.

\subsection{Directions in Time}

One acclaimed virtue of this version of the Brussels-Austin approach is the
ability of $\Lambda$ to provide time-asymmetry. The transformation $\Lambda$
is chosen so that time-asymmetry is guaranteed under its action. There exist,
nevertheless, two distinct transformations, $\Lambda_{+}$ and $\Lambda_{-}$,
corresponding to two distinct semigroups, $W_{t}^{+\ast}$ and $W_{t}^{-\ast}$,
respectively. $\Lambda_{+}$ corresponds to future-directed evolution toward
equilibrium along the positive $t$-axis and $\Lambda_{-}$ corresponds to
past-directed evolution toward equilibrium along the negative $t$-axis (Misra
and Prigogine 1983, p.422). This implies that there are two possible
probabilistic descriptions.

Why then do we not observe evolutions of the $W_{t}^{-\ast}$-type? To answer
this question the Brussels-Austin Group uses singular initial probability
distributions since nonsingular distributions can approach equilibrium in
either direction in time under $\Lambda_{+}$ and $\Lambda_{-}$. By translating
their conception of entropy into information-theoretic language, Courbage and
Prigogine (1983, p. 2414-5) showed that their formulation of the second law
requires infinite information for specifying the initial states of a singular
distribution evolving in the negative $t$-direction, but only finite
information for specifying the initial states for evolution in the positive
$t$-direction. This would render the initial conditions for systems to
approach equilibrium along the negative $t$-axis physically unrealizable: `Of
course, even a regular function close to a contracting fiber [$\Lambda_{-}%
$-type description] will require such a high information content that it will
be practically impossible to realize it for a given state of technology'
(Courbage and Prigogine 1983, p. 2416). Since singular probability
distributions are supposedly operationally unrealizable, they argue it is
physically impossible for unstable systems to evolve to equilibrium in the
negative $t$-direction. Hence their version of the second law acts as a
selection rule for initial states.

This argument is supposed to show why anti-thermodynamic behavior in the real
world is impossible (for a slightly different version, see Misra and Prigogine
1983). Nevertheless, the argument is problematic. The most fundamental
difficulty is that it conflates epistemic concepts (e.g. information,
empirical accessibility of states) with ontic concepts (e.g. actual states and
behaviors of systems).

Second, Courbage and Prigogine claimed that, `this selection rule expresses
the unrealizability of experiences in which a set of particles that undergo
several collisions will asymptotically emerge with parallel velocities' (1983,
p. 2413). As Sklar points out, however, spin-spin echo experiments represent
systems apparently exhibiting just the anti-thermodynamic behavior the
Brussels-Austin selection principle rules out (Sklar 1993, pp. 219-22). In
these experiments a number of molecules with a magnetic moment are initially
in a state where all the moments are aligned in the same direction. A
nonuniform magnetic field is then applied. In response to the field, the
magnetic moments begin to spin, but since the field is nonuniform, some
moments spin faster than others due to their spatial location with respect to
the field. After a time period $t_{a}$, the orientations of the magnetic
moments are completely random. At this point the nonuniform magnetic field is
reversed. The moments begin to spin in the opposite direction such that after
time $2t_{a}$, all the spins are aligned in the same initial direction. It
looks as if the system has been thermodynamically reversed.

The Brussels-Austin Group has a response to this objection. If the system is
left to itself, the orientations of the magnetic moments will continue to be
random and entropy continues to increase monotonically. If the magnetic field
is reversed at time $t_{a}$, the entropy of the system is decreased to a value
\textit{below} its initial value (it actually makes a discontinuous jump) due
to the external intervention of reversing the field (i.e. the system was
opened to an outside influence). As the moments reverse themselves, however,
the entropy continues its monotonic increase from its newly lowered value and
returns to its initial value at the point in time when the moments return to
their initial alignments.\footnote{A discussion of a generic time reversal
experiment in the context of a two state system (a ground state and an excited
state) is given in (Petrosky, Prigogine and Tasaki 1991, 200-202).}

A more fundamental problem with the selection principle argument is that it
turns on the definition of entropy. The conditions for the existence of the
microscopic entropy operator $M$ and, hence, for $\Lambda$, admit alternative
notions of entropy that have the opposite temporal behavior to the
Brussels-Austin Group definition ($\Lambda_{+}$ $=M_{+}^{%
%TCIMACRO{\U{bd}}%
%BeginExpansion
\frac12
%EndExpansion
}$). Recall that the microentropy operator $M$ is related to the existence of
a time operator $T$ for K-flows. There is a relationship between such time
operators and the Kolmogorov-Sinai entropy (Atmanspacher and Scheingraber
1987), and since K-flows are reversible, one could select an alternative
``entropy'' (e.g. characterized by negative rather than positive Lyapunov
exponents) with the opposite temporal direction endowing $T$ with the opposite
temporal direction and, then, construct an $M_{-}$. On this basis an argument
similar to that of Courbage and Prigogine could be formulated whose conclusion
is that the approach to equilibrium along the positive $t$-axis is
``impossible'' (e.g. Misra and Prigogine 1983, p. 427; Karakostas 1996, pp.
393-4). On what basis is one definition privileged over another? Why do we not
see about half of the systems approaching equilibrium in one time direction,
while the other half approach equilibrium in the opposite time direction?

The Brussels-Austin Group often responds to this type of objection by
appealing to experimental observations of time asymmetry. This amounts to
taking phenomenological laws as fundamental and thereby excludes all
definitions of entropy that licensed anti-thermodynamic behavior. Obviously
such a move comes at an explanatory cost. It is precisely these observations
that need explanation, but by taking them as fundamental the Brussels-Austin
Group gives up the ability to offer an explanation for the thermodynamic arrow
of time. In other words the acclaimed link between classical deterministic
systems and Markov processes, which was supposed to illuminate the mystery of
irreversibility, affords us no gain in understanding the puzzle of the second
law and is in danger of becoming circular.

\subsection{Problems with the ``Equivalence'' Thesis}

The ``equivalence'' between trajectory and probabilistic descriptions of
unstable system \textit{via} $\Lambda$ stands in need of further
clarification. If $\Lambda$ is a similarity transformation, it must preserve
the spacetime features of the physical system. In this approach, however, the
ontological elements of the two descriptions are supposed to be so different
(point states and trajectories vs. probability distributions) that the
implication should be that we have two different physical descriptions or
models of the system. Assume for the sake of argument that in unstable systems
smooth state space trajectories are physically irrelevant idealizations. This
view calls into question the validity of the classical deterministic
description which assumes such trajectories are physically meaningful.

Some insight into the ``equivalence'' thesis and the physical significance of
the similarity transformation can be found in the work by Gustafson and
colleagues (Gustafson and Goodrich 1980; Antoniou and Gustafson 1993;
Gustafson 1997; Antoniou, Gustafson and Suchanecki 1998). They have shown that
any Markovian semigroup dynamics arising from a coarse-grained projection of a
K-flow can be embedded into a larger Kolmogorov dynamical system. Moreover
many other kinds of Markovian semigroup dynamics can also be embedded into a
larger Kolmogorov system regardless of their origin (Gustafson 1997, pp.66-8
and references therein; Antoniou, Gustafson and Suchanecki 1998, pp. 114-8).
No specific results for embedding a Markovian dynamics induced by similarity
transformations exist at present as no concrete realizations of $\Lambda$ for
physical systems have been developed nor are many physical properties of such
transformations known (e.g. Antoniou, Gustafson and Suchanecki 1998, p. 119).
From this perspective, then, the equivalence of deterministic and
probabilistic descriptions \textit{via} $\Lambda$ needs further specification.
The physical significance of $\Lambda$ for unstable systems can be understood
minimally as a change of representation from the deterministic description to
a dynamics distinguishing stable and unstable manifolds of such systems.

In my judgment, the Brussels-Austin Group ought to have been arguing that the
probability description is the primary physical picture of the behavior of
unstable systems. After all, according to them the ontological elements of the
deterministic description are unrealizable and the probability
description--being irreducibly probabilistic--captures the dynamical behavior
of the probability distribution and the collective and long-range effects
within such systems that are missing from a trajectory description. In
addition it is precisely this irreducible probability that gives rise to the
claim that unstable classical systems can be intrinsically random or
indeterministic (Misra, Prigogine and Courbage 1979; Goldstein, Misra and
Courbage 1981). Finally Gustafson has demonstrated that the inverse
transformation $\Lambda^{-1}$ cannot be positivity preserving for K-flows, so
any reverse transformation (``embedding'') from the probabilistic description
induced by $\Lambda$ to a deterministic Kolmogorov dynamics must violate
positivity of probability (Gustafson 1997, pp. 61-2). So it appears that a
case can be made that under $\Lambda$, probabilistic models are more
physically accurate or appropriate for these systems. The similarity
transformation, then, should not be viewed as yielding an equivalence between
the two descriptions.\footnote{In more recent work, Prigogine and colleagues
explicitly argue that the trajectory description and the distribution
description are not equivalent for a class of models known as large
Poincar\'{e} systems (Petrosky and Prigogine 1996 and 1997; Part II).}

\section{Discussion}

Part of the explanation for the difficulties in the similarity transformation
approach lies in the fact that the Brussels-Austin Group started to work in an
inappropriate mathematical framework. They treated the operators $\Lambda$ and
$W_{t}^{\ast}$ as if they were defined on HS in the modified Koopman approach,
when what was actually needed was an extended space such as a rigged Hilbert
space (Bohm 1981, 2814; Obcemea and Br\"{a}ndas 1983; Petrosky and Prigogine
1996, pp. 481-2; Part II). Indeed Ord\'{o}\~{n}ez has recently shown that the
similarity transformation approach amounts to rigging a HS (1998). Furthermore
the fact that they were tacitly working in an extended space all along is
indicated by their interests in the evolution of distribution functions and
densities, the use of delta functions in the arguments on trajectories I
rehearsed above, as well as the presence of the semigroup operators
$W_{t}^{\ast}$ and the unbounded nature of the operators $U_{t}$ and
$W_{t}^{\ast}$. None of these elements are well defined on the whole of HS.

Although serious questions remain regarding how to demonstrate conclusively
that the past-directed semigroups are somehow unphysical and regarding the
mathematical difficulty of the similarity transformation approach, this
earlier attempt by the Brussels-Austin Group to develop a new approach to
thermodynamics and SM does achieve some milestones. It provides a generalized
formalism allowing a unified mathematical treatment of deterministic and
probabilistic systems and has potential application to QM systems as well,
holding out the possibility of a unified treatment for statistical physics at
both the microphysical and macrophysical levels. In addition the framework
appears equally suitable to QM and CM suggesting that the two levels of
description may themselves become unified. Provided the transformation
$\Lambda$ can be found for the unstable SM systems in question, the
Brussels-Austin Group demonstrated a \textit{mathematical} relation between
the time-reversible dynamics of the trajectory description and the
time-irreversible dynamics of the Markov description. This suggests that if
they could produce an alternative justification for taking the Markov
description as primary, they may still be able to reconcile the irreversible
behavior with the standard time-reversible SM. Along the way there were new
mathematical developments, particularly in the area of time operators.

The realization that this formalism was tacitly assuming an extension beyond
HS was one of the reasons that motivated the Brussels-Austin Group to move to
rigged Hilbert spaces. Another motivation for switching formalisms was that
realistic physical models become almost mathematically intractable in the
similarity transformation approach. Results were only obtained for discrete
mathematical systems like the Baker's transformation, never for continuous
mathematical systems much less realistic physical models. The more recent work
in rigged Hilbert spaces holds forth the promise of overcoming many of these
difficulties, so I will discuss it in Part II.

\bigskip

\textbf{References}

Antoniou, I. and Gustafson, K. (1993) `From Probabilistic Descriptions to
Deterministic Dyanmics', \textit{Physica A} \textbf{197}, 153-66.

Antoniou, I., Gustafson, K. and Suchanecki, Z. (1998) `From Stochastic
Semigroups to Chaotic Dynamics', in A. Bohm, H-D. Doebner and P. Kielanowski
(eds.), \textit{Irreversibility and Causality: Semigroups and Rigged Hilbert
Spaces} (Berlin: Springer-Verlag).

Antoniou, I., Prigogine, I., Sadovnichii, V. and Shkarin, S. (2000) `Time
Operator for Diffusion', \textit{Chaos, Solitons and Fractals} \textbf{11}, 465-77.

Antoniou, I., Sadovnichii, V. and Shkarin, S. (1999) `Time Operators and Shift
Representation of Dynamical Systems', \textit{Physica A} \textbf{269}, 299-313.

Antoniou, I. and Suchanecki, Z. (2000) `Non-uniform Time Operator, Chaos and
Wavelets on the Interval', \textit{Chaos, Solitons and Fractals} \textbf{11}, 423-35.

Atmanspacher, H., Bishop, R. and Amann, A. (2002) `Extrinsic and Intrinsic
Irreversibility in Probabilistic Dynamical Laws', in A. Khrennikov (ed),
\textit{Quantum Probability and White Noise Analysis Volume XIII} (Singapore:
World Scientific).

Atmanspacher, H. and Scheingraber H. (1987) `A Fundamental Link between System
Theory and Statistical Mechanics', \textit{Foundations of Physics}
\textbf{17}, 939-63.

Batterman, R. (1991) `Randomness and Probability in Dynamical Theories: On the
Proposals of the Prigogine School', \textit{Philosophy of Science}
\textbf{58}, 241-263.

Bishop, R. (forthcoming) `On Separating Predictability and Determinism',
\textit{Erkenntnis}.

Bohm, A. (1981) `Resonance Poles and Gamow Vectors in the Rigged Hilbert Space
Formulation of Quantum Mechanics', \textit{Journal of Mathematical Physics}
\textbf{22}, 2813-22.

Braunss, G. (1984) `On the Construction of State Spaces for Classical
Dynamical Systems with a Time-dependent Hamiltonian Function', \textit{Journal
of Mathematical Physics} \textbf{25}, 266-70.

\_\_\_\_\_\_\_\_\_ (1985) `Intrinsic Stochasticity of Dynamical Systems',
\textit{Acta Applicandae Mathematicae} \textbf{3}, 1-21.

Bricmont, J. (1995) `Science of Chaos or Chaos in Science?' \textit{Physicalia
Magazine} \textbf{17}, 159-208.

Courbage, M. and Misra, B. (1980), `On the Equivalence between Bernoulli
Dynamical Systems and Stochastic Markov Processes', \textit{Physica}
\textbf{104A}, 359-77.

Courbage, M., and Prigogine, I. (1983) `Intrinsic Randomness and Intrinsic
Irreversibility in Classical Dynamical Systems', \textit{Proceedings of the
National Academy of Sciences USA} \textbf{80}, 2412-2416.

Dougherty, J. (1993) `Explaining Statistical Mechanics', \textit{Studies in
History and Philosophy of Science} \textbf{24}: 843-66.

George, C. (1973a) `Subdynamics and Correlations', \textit{Physica}
\textbf{65}, 277-302.

\_\_\_\_\_\_\_\_ (1973b) \textit{Lectures in Statistical Physics II, Lecture
Notes in Physics} (Berlin: Springer-Verlag).

Goldstein, S., Misra, B, and Courbage, M. (1981) `On Intrinsic Randomness of
Dynamical Systems', \textit{Journal of Statistical Physics} \textbf{25}, 111-126.

Goodrich, K, Gustafson, K and Misra, B. (1980) `A Converse to Koopman's
Lemma', \textit{Physica 102A}, 379-88.

Gustafson, K. (1997) \textit{Lectures on Computational Fluid Dynamics,
Mathematical Physics, and Linear Algebra} (Singapore: World Scientific).

Gustafson, K. and Goodrich, K. (1980) `A Banach-Lamperti Theorem and
Similarity Transformations in Statistical Mechanics', \textit{Colloquium of
the Mathematical Society of Janos Bolyai} \textbf{35}, 567-79.

Hale, J and Ko\c{c}ak, H. (1991) \textit{Dynamics and Bifurcations} (Berlin: Springer-Verlag).

Karakostas, V. (1996) `On the Brussels School's Arrow of Time in Quantum
Theory', \textit{Philosophy of Science} \textbf{63}, 374-400.

Koopman, B. (1931) `Hamiltonian Systems and Transformations in Hilbert Space',
\textit{Proceedings of the National Academy of Sciences }\textbf{17}, 315-18.

Mercik, S. And Weron, K. (2000) `Application of the Internal Time Operators
for the Renyi Map', \textit{Chaos, Solitons and Fractals} \textbf{11}, 437-42.

Misra, B. (1978) `Nonequilibrium Entropy, Lyapunov Variables, and Ergodic
Properties of Classical Systems', \textit{Proceedings of the National Academy
of Sciences} \textbf{75}, 1627-31.

Misra, B. and Prigogine, I. (1983) `Irreversibility and Nonlocality',
\textit{Letters in Mathematical Physics} \textbf{7}, 421-429.

Misra, B., Prigogine, I. and Courbage, M. (1979) `From Deterministic Dynamics
to Probabilistic Descriptions', \textit{Physica A} \textbf{98A}, 1-26.

Nicolis, G. and Prigogine, I. (1989) \textit{Exploring Complexity: An
Introduction} (New York: W. H. Freeman and Company).

Obcemea, Ch. and Br\"{a}ndas, E. (1983) `Analysis of Prigogine's Theory of
Subdynamics', \textit{Annals of Physics} \textbf{151}, 383-430.

Ord\'{o}\~{n}ez, A. (1998) `Rigged Hilbert Spaces Associated with
Misra-Prigogine-Courbage Theory of Irreversibility', \textit{Physica A}
\textbf{252}, 362-76.

Petrosky, T. and Prigogine, I. (1996) `Poincar\'{e} Resonances and the
Extension of Classical Dynamics', \textit{Chaos, Solitons \& Fractals}
\textbf{7}: 441-497.

\_\_\_\_\_\_\_\_\_\_\_\_\_\_\_\_\_ (1997) `The Extension of Classical Dynamics
for Unstable Hamiltonian Systems', \textit{Computers \& Mathematics with
Applications} \textbf{34}: 1-44.

Petrosky, T., Prigogine, I. and Tasaki, S. (1991) `Quantum Theory of
Non-integrable Systems', \textit{Physica A} \textbf{175}, 175-242.

Prigogine, I. (1962) \textit{Non-Equilibrium Statistical Mechanics} (New York:
Interscience Publishers).

Prigogine, I., George, C., Henin, F. (1969) `Dynamical and Statistical
Descriptions of N-Body Systems', \textit{Physica} \textbf{45}, 418-34.

Prigogine, I., George, C., Henin, F. and Rosenfeld, L. (1973) `A Unified
formulation of Dynamics and Thermodynamics', \textit{Chemica Scripta}
\textbf{4}, 5-32.

Schwartz, L. (1950, 1951) \textit{Th\'{e}orie des distributions I, II} (Paris: Hermann).

Sklar, L. (1993) \textit{Physics and Chance} (Cambridge: Cambridge University Press).

Suchanecki, Z. (1992), `On Lambda and Internal Time Operators' \textit{Physica
A} \textbf{187}, 249-66.

Suchanecki, Z. Antoniou, I. and Tasaki, S. (1994) `Nonlocality of the
Misra-Prigogine-Courbage Semigroup', \textit{Journal of Statistical Physics}
\textbf{75}, 919-28.
\end{subequations}
\end{document}